# Topology Discovery Using Cisco Discovery Protocol

**Sergio R. Rodriguez**
School of Computer Science and Engineering
The University of New South Wales
Sydney, Australia

*Abstract*

In this paper we address the problem of discovering network topology in proprietary networks. Namely, we investigate topology discovery in Cisco-based networks. Cisco devices run Cisco Discovery Protocol (CDP) which holds information about these devices. We first compare properties of topologies that can be obtained from networks deploying CDP versus Spanning Tree Protocol (STP) and Management Information Base (MIB) Forwarding Database (FDB). Then we describe a method of discovering topology of CDP-based networks. Our experiments show that the physical topology of the network including links that are in Forwarding Block state can be discovered.

## 1. Introduction

Meeting the increasingly demanding Service Level Agreement (SLA) requirements for providing quality of service in network-based applications requires accurate and improved network management visibility and accuracy. Providing stable and efficient end-to-end application performance management solutions requires accurate knowledge of the involving network infrastructure.

Thus, network topologies mostly interest network engineers and managers since it is part of their job to handle tasks that are directly or indirectly related to network topology. Handling tasks and issues related or caused by changes in network topology requires knowledge of the network connectivity infrastructure. For instance, finding the root cause of a malfunction caused by network user, software, or device requires complete knowledge of the network topology. Besides troubleshooting, network planning and extending require clear understanding of the constant changing topology. In general, network operators and managers should have a detailed view of the current state of the network including devices and links at any time.

Cisco Discovery Protocol is a network management protocol used to manage Cisco based devices [6]. The advantage of using CDP to discover the topology is that CDP provides detailed information about the devices' connectivity. This method is useful for smaller Cisco-based or networks that deploy devices that are consistent with CDP.

In this paper we present Cisco Discovery Protocol and describe the method we developed to discover the data link network topology based on CDP.

## 2. Topology Discovery Protocols and Methods

In this section we first give some background about Cisco Discovery Protocol then we describe some other vendor independent tools and methods.

### 1) Cisco Discovery Protocol

Cisco Discovery Protocol is a proprietary data link layer network protocol developed by Cisco Systems [6]. CDP runs on Cisco devices such as access servers, routers, bridges and switches and it maintains information about network equipment status.

One of the key uses of CDP is to obtain information about neighboring devices and discover the network devices' interconnectivity, which can be extracted from CISCO-CDP-MIB object. Information obtained from both CDP and SNMP [10] can be combined to learn the neighboring devices of a given network device. The addresses of the newly discovered devices can be queried recursively to obtain the full topology of the network.

Besides connectivity information, CDP sends multicast messages periodically about the current status of the device and configurations. Each device also listens to messages sent by other devices to determine the status of other devices' interfaces.

### 2) Vendor-Independent Topology Discovery Methods

Even though the Internet Engineering Task Force (IETF) has proposed a common standard for Data Link topology discovery, it may take years until it becomes a standard and gets deployed in network devices.

Current research has focused mainly on defining Simple Network Management Protocol-Management Information Base (SNMP-MIB) which holds network-topology data. The Physical topology MIB, known as Request for comment (RFC) 2922 [7], defines how physical topology information will be maintained.

Many researchers have addressed topology discovery in heterogeneous networks. The article "Topology

Discovery in Heterogeneous IP Networks" [3] describes a method for data link topology discovery when MIB holds complete information about forwarding databases.

In the paper "Topology Discovery for Large Ethernet Network" [8], the authors describe the "minimum knowledge required" to find the connectivity between two interfaces.

The articles "Ethernet Topology Discovery for Networks with Incomplete Information" [4] and "Discovering Network Topology of Large Multisubnet Ethernet Networks" [5] propose techniques for data link topology discovery in the absence of dumb devices when the devices don't hold complete forwarding database, and when the presence of dumb devices but network devices hold complete forwarding information, respectively.

Black et al [2] describe in the paper "Ethernet Topology Discovery without Network Assistance" a different method for topology discovery based on packets that are being injected in the network from hosts. These packets hold information about the network routes, which is aggregated to infer the topology.

Finally, in the paper "Taking the Skeletons Out of the Closets: A Simple and Efficient Topology Discovery Scheme for Large Ethernet LANs" [1], Bejerano presents a method to infer the topology in the presence of dumb devices when forwarding tables are complete and the network has an anchor nodes; a node that is contained in every devices forwarding table. Table 1 presents the differences between various topology discovery methods.

| Topology Discovery Method | CDP | STP | FDB |
|---|---|---|---|
| Active Topology | Yes | Yes | No |
| None Active Topology | Yes | Yes | No |
| Router/Switch Discovery | Yes | No | Yes |
| Host Discovery | No | No | Yes |
| Shared Segment Discovery | No | No | Yes |

**Table 1:** Comparison between features supported by Data Link topology discovery methods.

### 3. Topology Discovery Algorithm

This section describes the details the design of the algorithm we designed to discover the topology for networks deploying CDP and the experiments we ran to discover the topology.

*1) Experiment Design and Execution*

Our topology discovery process starts by taking the IP address of a root device. The root device can be any device in the network that its topology being discovered. The algorithm reads the neighboring devices' IP address from the Cisco MIB, puts them a queue, and repeats the same process for each IP address in the queue until the address queue is empty. The root is considered as level 1 and devices that are connected to it are level 2 and so on.

We describe our topology discovery method in the network depicted in Figure 1. The root is the switch with IP address 192.168.10.1. Table 2 shows the network devices retrieved at each iteration of the topology discovery procedure.

We ran the algorithm on a network that consists of 92 Cisco devices and read the Cisco MIB data using SNMP WALK and GETBULK commands [9].

*1) Results*

Our method retrieved the data from all devices in about 5 minutes and the topology discovery process took less than 10 seconds. Our method didn't discover only the active topology, but all connections between devices including the currently blocked ones by the Spanning Tree Protocol. However, links that were connected between interfaces which were administratively blocked couldn't be discovered.

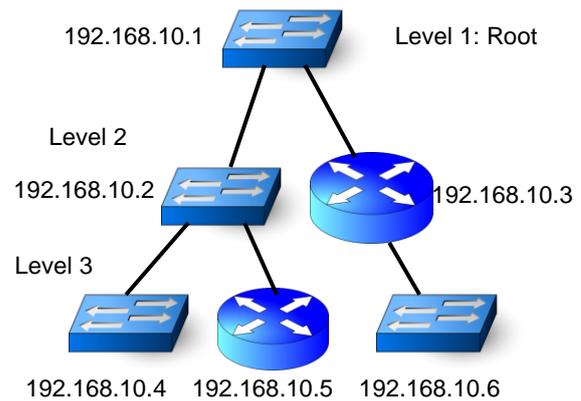

**Figure 1:** Topology discover model.

| Node Discovery Step | Device IP | Neighbors |
|---|---|---|
| 1 | 192.168.10.1 | 192.168.10.2 |
|  |  | 192.168.10.3 |
| 2 | 192.168.10.2 | 192.168.10.4 |
|  |  | 192.168.10.5 |
|  | 192.168.10.3 | 192.168.10.6 |
| 3 | 192.168.10.4 | 192.168.10.4 |
|  | 192.168.10.5 | 192.168.10.5 |
|  | 192.168.10.6 | 192.168.10.6 |

**Table 2:** Topology discovery process for the network depicted in Figure 1.

### 4. Conclusion

Within this paper we have explained the differences between various topology discovery models. Then we have presented a method for Data Link topology discovery from the vendor implemented protocol (CDP). The method has been implemented and example has been presented.